# InterCloud: Utility-Oriented Federation of Cloud Computing Environments for Scaling of Application Services

Rajkumar Buyya[1, 2], Rajiv Ranjan[3], Rodrigo N. Calheiros[1]

[1] **Clou**d Computing and **D**istributed **S**ystems (CLOUDS) Laboratory
Department of Computer Science and Software Engineering
The University of Melbourne, Australia

[2] Manjrasoft Pty Ltd, Australia

[3] School of Computer Science and Engineering
The University of New South Wales, Sydney, Australia

## Abstract

Cloud computing providers have setup several data centers at different geographical locations over the Internet in order to optimally serve needs of their customers around the world. However, existing systems do not support mechanisms and policies for dynamically coordinating load distribution among different Cloud-based data centers in order to determine optimal location for hosting application services to achieve reasonable QoS levels. Further, the Cloud computing providers are unable to predict geographic distribution of users consuming their services, hence the load coordination must happen automatically, and distribution of services must change in response to changes in the load. To counter this problem, we advocate creation of federated Cloud computing environment (InterCloud) that facilitates just-in-time, opportunistic, and scalable provisioning of application services, consistently achieving QoS targets under variable workload, resource and network conditions. The overall goal is to create a computing environment that supports dynamic expansion or contraction of capabilities (VMs, services, storage, and database) for handling sudden variations in service demands.

This paper presents vision, challenges, and architectural elements of InterCloud for utility-oriented federation of Cloud computing environments. The proposed InterCloud environment supports scaling of applications across multiple vendor clouds. We have validated our approach by conducting a set of rigorous performance evaluation study using the CloudSim toolkit. The results demonstrate that federated Cloud computing model has immense potential as it offers significant performance gains as regards to response time and cost saving under dynamic workload scenarios.

# 1. Introduction

In 1969, Leonard Kleinrock [1], one of the chief scientists of the original Advanced Research Projects Agency Network (ARPANET) project which seeded the Internet, said: *"As of now, computer networks are still in their infancy, but as they grow up and become sophisticated, we will probably see the spread of 'computer utilities' which, like present electric and telephone utilities, will service individual homes and offices across the country."* This vision of computing utilities based on a service provisioning model anticipated the massive transformation of the entire computing industry in the 21$^{st}$ century whereby computing services will be readily available on demand, like other utility services available in today's society. Similarly, computing service users (consumers) need to pay providers only when they access computing services. In addition, consumers no longer need to invest heavily or encounter difficulties in building and maintaining complex IT infrastructure.

In such a model, users access services based on their requirements without regard to where the services are hosted. This model has been referred to as *utility computing*, or recently as *Cloud computing* [3][7]. The latter term denotes the infrastructure as a "Cloud" from which businesses and users are able to access application services from anywhere in the world on demand. Hence, Cloud computing can be classified as a new paradigm for the dynamic provisioning of computing services, typically supported by state-of-the-art data centers containing ensembles of networked Virtual Machines.

Cloud computing delivers infrastructure, platform, and software (application) as services, which are made available as subscription-based services in a pay-as-you-go model to consumers. These services in industry are respectively referred to as Infrastructure as a Service (IaaS), Platform as a Service (PaaS), and Software as a Service (SaaS). A Berkeley Report in Feb 2009 states "Cloud computing, the long-held dream of computing as a utility, has the potential to transform a large part of the IT industry, making software even more attractive as a service" [2].

Clouds aim to power the next generation data centers by architecting them as a network of virtual services (hardware, database, user-interface, application logic) so that users are able to access and deploy applications from anywhere in the world on demand at competitive costs depending on users QoS (Quality of Service) requirements [3]. Developers with innovative ideas for new Internet services no longer require large capital outlays in hardware to deploy their service or human expense to operate it [2]. It offers significant benefit to IT companies by freeing them from the low level task of setting up basic hardware (servers) and software infrastructures and thus enabling more focus on innovation and creating business value for their services.

The business potential of Cloud computing is recognised by several market research firms including IDC, which reports that worldwide spending on Cloud services will grow from $16 billion by 2008 to $42 billion in 2012. Furthermore, many applications making use of these utility-oriented computing systems such as clouds emerge simply as catalysts or market makers that bring buyers and sellers together. This creates several trillion dollars of worth to the utility/pervasive com-puting industry as noted by Sun Microsystems co-founder Bill Joy [4]. He also indicated "It would take time until these markets to mature to generate this kind of value. Predicting now which companies will capture the value is impossible. Many of them have not even been created yet."

## 1.1 Application Scaling and Cloud Infrastructure: Challenges and Requirements

Providers such as Amazon [15], Google [16], Salesforce [21], IBM, Microsoft [17], and Sun Microsystems have begun to establish new data centers for hosting Cloud computing application services such as social networking and gaming portals, business applications (e.g., SalesForce.com), media content delivery, and scientific workflows. Actual usage patterns of many real-world application services vary with time, most of the time in unpredictable ways. To illustrate this, let us consider an "elastic" application in the business/social networking domain that needs to scale up and down over the course of its deployment.

### Social Networking Web Applications

Social networks such as Facebook and MySpace are popular Web 2.0 based applications. They serve dynamic content to millions of users, whose access and interaction patterns are hard to predict. In addition, their features are very dynamic in the sense that new plug-ins can be created by independent developers, added to the main system and used by other users. In several situations load spikes can take place, for instance, whenever new system features become popular or a new plug-in application is deployed. As these social networks are organized in communities of highly interacting users distributed all over the world, load spikes can take place at different locations at any time. In order to handle unpredictable seasonal and geographical changes in system workload, an automatic scaling scheme is paramount to keep QoS and resource consumption at suitable levels.

Social networking websites are built using multi-tiered web technologies, which consist of application servers such as IBM WebSphere and persistency layers such as the MySQL relational database. Usually, each component runs in a separate virtual machine, which can be hosted in data centers that are owned by different cloud computing providers. Additionally, each plug-in developer has the freedom to choose which Cloud computing provider offers the services that are more suitable to run his/her plug-in. As a consequence, a typical social networking web application is formed by hundreds of different services, which may be hosted by dozens of Cloud data centers around the world. Whenever there is a variation in temporal and spatial locality of workload, each application component must dynamically scale to offer good quality of experience to users.

## 1.2 Federated Cloud Infrastructures for Elastic Applications

In order to support a large number of application service consumers from around the world, Cloud infrastructure providers (i.e., IaaS providers) have established data centers in multiple geographical locations to provide redundancy and ensure reliability in case of site failures. For example, Amazon has data centers in the US (e.g., one in the East Coast and another in the West Coast) and Europe. However, currently they (1) expect their Cloud customers (i.e., SaaS providers) to express a preference about the location where they want their application services to be hosted and (2) don't provide seamless/automatic mechanisms for scaling their hosted services across multiple, geographically distributed data centers. This approach has many shortcomings, which include (1) it is difficult for Cloud customers to determine in advance the best location for hosting their services as they may not know origin of consumers of their services and (2) Cloud SaaS providers may not be able to meet QoS expectations of their service consumers originating from multiple geographical locations. This necessitates building mechanisms for seamless federation of data centers of a Cloud provider or providers supporting dynamic scaling of applications across multiple domains in order to meet QoS targets of Cloud customers.

In addition, no single Cloud infrastructure provider will be able to establish their data centers at all possible locations throughout the world. As a result Cloud application service (SaaS) providers will have difficulty in meeting QoS expectations for all their consumers. Hence, they would like to make use of services of multiple Cloud infrastructure service providers who can provide better support for their specific consumer needs. This kind of requirements often arises in enterprises with global operations and applications such as Internet service, media hosting, and Web 2.0 applications. This necessitates building mechanisms for federation of Cloud infrastructure service providers for seamless provisioning of services across different Cloud providers. There are many challenges involved in creating such Cloud interconnections through federation.

To meet these requirements, next generation Cloud service providers **should be able** to: (i) dynamically expand or resize their provisioning capability based on sudden spikes in workload demands by leasing available computational and storage capabilities from other Cloud service providers; (ii) operate as parts of a market driven resource leasing federation, where application service providers such as Salesforce.com host their services based on negotiated Service Level Agreement (SLA) contracts driven by competitive market prices; and (iii) deliver on demand, reliable, cost-effective, and QoS aware services based on virtualization technologies while ensuring high QoS standards and minimizing service costs. They need to be able to utilize market-based utility models as the basis for provisioning of virtualized software services and federated hardware infrastructure among users with heterogeneous applications and QoS targets.

## 1.3 Research Issues

The diversity and flexibility of the functionalities (dynamically shrinking and growing computing systems) envisioned by federated Cloud computing model, combined with the magnitudes and uncertainties of its components (workload, compute servers, services, workload), pose difficult problems in effective provisioning and delivery of application services. Provisioning means "high-level management of computing, network, and storage resources that allow them to effectively provide and deliver services to customers". In particular, finding efficient solutions for following challenges is critical to exploiting the potential of federated Cloud infrastructures:

**Application Service Behavior Prediction:** It is critical that the system is able to predict the demands and behaviors of the hosted services, so that it intelligently undertake decisions related to dynamic scaling or de-scaling of services over federated Cloud infrastructures. Concrete prediction or forecasting models must be built before the behavior of a service, in terms of computing, storage, and network bandwidth requirements, can be predicted accurately. *The real challenge in devising such models is accurately learning and fitting statistical functions [31] to the observed distributions of service behaviors such as request arrival pattern, service time distributions, I/O system behaviors, and network usage.* This challenge is further aggravated by the existence of statistical correlation (such as stationary, short- and long-range dependence, and pseudo-periodicity) between different behaviors of a service.

**Flexible Mapping of Services to Resources:** With increased operating costs and energy requirements of composite systems, it becomes critical to maximize their efficiency, cost-effectiveness, and utilization [30] . The process of mapping services to resources is a complex undertaking, as it requires the system to *compute the best software and hardware configuration (system size and mix of resources) to ensure that QoS targets of services are achieved, while maximizing system efficiency and utilization. This process is further complicated by the uncertain behavior of resources and services.* Consequently, there is an immediate need to devise performance modeling and market-based service mapping techniques that ensure efficient system utilization without having an unacceptable impact on QoS targets.

**Economic Models Driven Optimization Techniques:** The market-driven decision making problem [6] is a combinatorial optimization problem that searches the optimal combinations of services and their deployment plans. Unlike many existing multi-objective optimization solutions, the optimization models that ultimately aim to optimize both resource-centric (utilization, availability, reliability, incentive) and user-centric (response time, budget spent, fairness) QoS targets need to be developed.

**Integration and Interoperability:** For many SMEs, there is a large amount of IT assets in house, in the form of line of business applications that are unlikely to ever be migrated to the cloud. Further, there is huge amount of sensitive data in an enterprise, which is unlikely to migrate to the cloud due to privacy and security issues. As a result, there is a need to look into issues related to integration and interoperability between the software on premises and the services in the cloud. In particular [28]: (i) Identity management: authentication and authorization of service users; provisioning user access; federated security model; (ii) Data Management: not all data will be stored in a relational database in the cloud, eventual consistency (BASE) is taking over from the traditional ACID transaction guarantees, in order to ensure sharable data structures that achieve high scalability. (iii) Business process orchestration: how does integration at a business process level happen across the software on premises and service in the

Cloud boundary? Where do we store business rules that govern the business process orchestration?

**Scalable Monitoring of System Components:** Although the components that contribute to a federated system may be distributed, existing techniques usually employ centralized approaches to overall system monitoring and management. We claim that centralized approaches are not an appropriate solution for this purpose, due to concerns of scalability, performance, and reliability arising from the management of multiple service queues and the expected large volume of service requests. *Monitoring of system components is required for effecting on-line control through a collection of system performance characteristics.* Therefore, we advocate architecting service monitoring and management services based on decentralized messaging and indexing models [27].

### 1.4 Overall Vision

To meet aforementioned requirements of auto-scaling Cloud applications, future efforts should focus on design, development, and implementation of software systems and policies for federation of Clouds across network and administrative boundaries. The key elements for enabling federation of Clouds and auto-scaling application services are: Cloud Coordinators, Brokers, and an Exchange. The resource provisioning within these federated clouds will be driven by market-oriented principles for efficient resource allocation depending on user QoS targets and workload demand patterns. To reduce power consumption cost and improve service localization while complying with the Service Level Agreement (SLA) contracts, new on-line algorithms for energy-aware placement and live migration of virtual machines between Clouds would need to be developed. The approach for realisation of this research vision consists of investigation, design, and development of the following:

- Architectural framework and principles for the development of utility-oriented clouds and their federation
- A Cloud Coordinator for exporting Cloud services and their management driven by market-based trading and negotiation protocols for optimal QoS delivery at minimal cost and energy.
- A Cloud Broker responsible for mediating between service consumers and Cloud coordinators.
- A Cloud Exchange acts as a market maker enabling capability sharing across multiple Cloud domains through its match making services.
- A software platform implementing Cloud Coordinator, Broker, and Exchange for federation.

The rest of this paper is organized as follows: First, a concise survey on the existing state-of-the-art in Cloud provisioning is presented. Next, the comprehensive description related to overall system architecture and its elements that forms the basis for constructing federated Cloud infrastructures is given. This is followed by some initial experiments and results, which quantifies the performance gains de-livered by the proposed approach. Finally, the paper ends with brief conclusive remarks and discussion on perspective future research directions.

## 2. State-of-the-art in Cloud Provisioning

The key Cloud platforms in Cloud computing domain including Amazon Web Services [15], Microsoft Azure [17], Google AppEngine [16], Manjrasoft Aneka [32], Eucalyptus [22], and GoGrid [23] offer a variety of pre-packaged services for monitoring, managing and provisioning resources and application services. However, the techniques implemented in each of these Cloud platforms vary (refer to Table 1).

The three Amazon Web Services (AWS), Elastic Load Balancer [25], Auto Scaling and CloudWatch [24] together expose functionalities which are required for undertaking provisioning of application services on Amazon EC2. Elastic Load Balancer service automatically provisions incoming application workload across available Amazon EC2 instances. Auto-Scaling service can be used for dynamically scaling-in or scaling-out the number of Amazon EC2 instances for handling changes in service demand patterns. And finally the CloudWatch service can be integrated with above services for strategic decision making based on real-time aggregated resource and service performance information.

Table 1: Summary of provisioning capabilities exposed by public Cloud platforms

| Cloud Platforms | Load Balancing | Provisioning | Auto Scaling |
|---|---|---|---|
| Amazon Elastic Compute Cloud | √ | √ | √ |
| Eucalyptus | √ | √ | × |
| Microsoft Windows Azure | √ | √ (fixed templates so far) | √ (Manual) |
| Google App Engine | √ | √ | √ |
| Manjrasoft Aneka | √ | √ | √ |
| GoGrid Cloud Hosting | √ | √ | √ (Programmatic way only) |

Manjrasoft Aneka is a platform for building and deploying distributed applications on Clouds. It provides a rich set of APIs for transparently exploiting distributed resources and expressing the business logic of applications by using the preferred programming abstractions. Aneka is also a market-oriented Cloud platform since it allows users to build and schedule applications, provision resources and monitor results using pricing, accounting, and QoS/SLA services in private and/or public (leased) Cloud environments. Aneka also allows users to build different run-time environments such as enterprise/private Cloud by harness computing re-sources in network or enterprise data centers, public Clouds such as Amazon EC2, and hybrid clouds by combining enterprise private Clouds managed by Aneka with resources from Amazon EC2 or other enterprise Clouds build and managed using technologies such as XenServer.

Eucalyptus [22] is an open source Cloud computing platform. It is composed of three controllers. Among the controllers, the Cluster Controller is a key component to application service provisioning and load balancing. Each Cluster Controller is hosted on the head node of a cluster to interconnect outer public networks and inner private networks together. By monitoring the state information of instances in the pool of server controllers, the Cluster Controller can select the available service/server for provisioning incoming requests. However, as compared to AWS, Eucalyptus still lacks some of the critical functionalities, such as auto scaling for built-in provisioner.

Fundamentally, Windows Azure Fabric [17] has a weave-like structure, which is composed of node (servers and load balancers), and edges (power, Ethernet and serial communications). The Fabric Controller manages a service node through a built-in service, named Azure Fabric Controller Agent, which runs in the background, tracking the state of the server, and reporting these metrics to the Controller. If a fault state is reported, the Controller can manage a reboot of the server or a migration of services from the current server to other healthy servers. Moreover, the Controller also supports service provisioning by matching the services against the VMs that meet required demands.

GoGrid Cloud Hosting offers developers the F5 Load Balancers [23] for distributing application service traffic across servers, as long as IPs and specific ports of these servers are attached. The load balancer allows Round Robin algorithm and Least Connect algorithm for routing application service requests. Also, the load balancer is able to sense a crash of the server, redirecting further requests to other available servers. But currently, GoGrid Cloud Hosting only gives developers programmatic APIs to implement their custom auto-scaling service.

Unlike other Cloud platforms, Google App Engine offers developers a scalable platform in which applications can run, rather than providing access directly to a customized virtual machine. Therefore, access to the underlying operating system is restricted in App Engine. And load-balancing strategies, service provisioning and auto scaling are all automatically managed by the system behind the scenes. However, at this time Google App Engine can only support provisioning of web hosting type of applications.

However, no single Cloud infrastructure providers have their data centers at all possible locations throughout the world. As a result Cloud application service (SaaS) providers will have difficulty in meeting QoS expectations for all their users. Hence, they would prefer to logically construct federated Cloud infrastructures (mixing multiple public and private clouds) to provide better support for their specific user needs. This kind of requirements often arises in enterprises with global operations and applications such as Internet service, media hosting, and Web 2.0 applications. This necessitates building technologies and algorithms for seamless federation of Cloud infrastructure service providers for autonomic provisioning of services across different Cloud providers.

## 3. System Architecture and Elements of InterCloud

Figure 1 shows the high level components of the service-oriented architectural framework consisting of client's brokering and coordinator services that support utility-driven federation of clouds: application scheduling, resource allocation and migration of workloads. The architecture cohesively couples the administratively and topologically distributed storage and computes capabilities of Clouds as parts of single resource leasing abstraction. The system will ease the cross-domain capabilities integration for on demand, flexible, energy-efficient, and reliable access to the infrastructure based on emerging virtualization technologies [8][9].

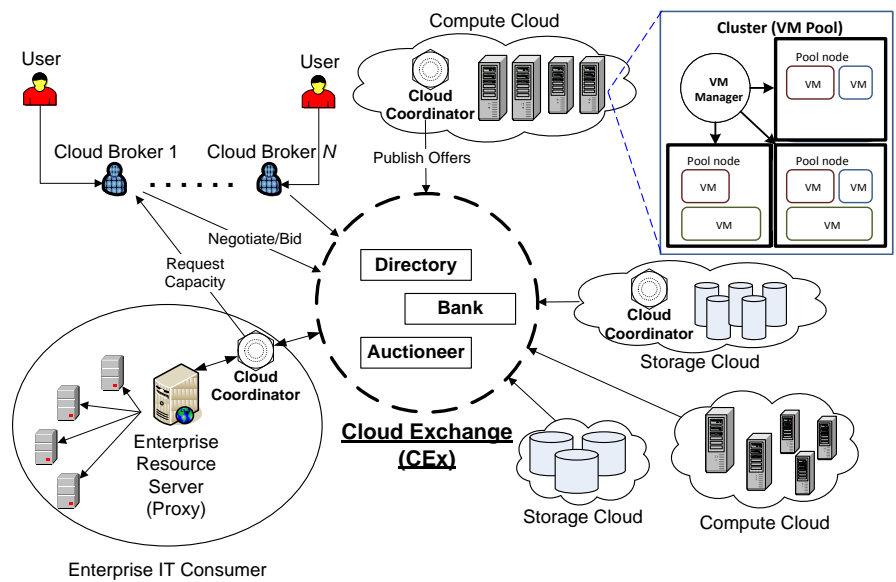

Figure 1: Federated network of clouds mediated by a Cloud exchange.

The Cloud Exchange (CEx) acts as a market maker for bringing together service producers and consumers. It aggregates the infrastructure demands from the application brokers and evaluates them against the available supply currently published by the Cloud Coordinators. It supports trading of Cloud services based on competitive economic models [6] such as commodity markets and auctions. CEx allows the participants (Cloud Coordinators and Cloud Brokers) to locate providers and consumers with fitting offers. Such markets enable services to be commoditized and thus, would pave the way for creation of dynamic market infrastructure for trading based on SLAs. An SLA specifies the details of the service to be provided in terms of metrics agreed upon by all parties, and incentives and penalties for meeting and violating the expectations, respectively. The availability of a banking system within the market ensures that financial transactions pertaining to SLAs between participants are carried out in a secure and dependable environ-ment. Every client in the federated platform needs to instantiate a Cloud Brokering service that can dynamically establish service contracts with Cloud Coordinators via the trading functions exposed by the Cloud Exchange.

### 3.1 Cloud Coordinator (CC)

The Cloud Coordinator service is responsible for the management of domain specific enterprise Clouds and their membership to the overall federation driven by market-based trading and negotiation protocols. It provides a programming, management, and deployment environment for applications in a federation of Clouds. Figure 2 shows a detailed depiction of resource management components in the Cloud Coordinator service.

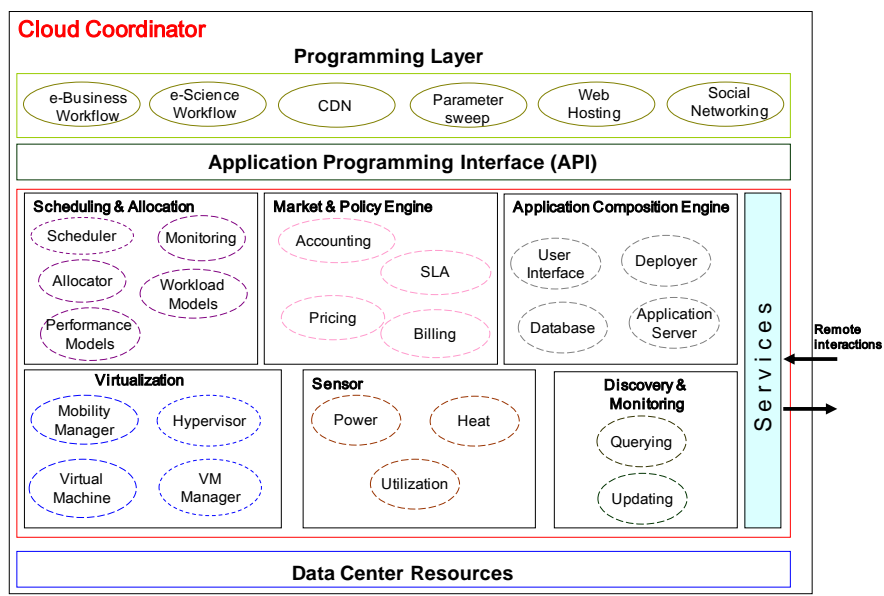

Figure 2. Cloud Coordinator software architecture.

The Cloud Coordinator exports the services of a cloud to the federation by implementing basic functionalities for resource management such as scheduling, allocation, (workload and performance) models, market enabling, virtualization, dynamic sensing/monitoring, discovery, and application composition as discussed below:

**Scheduling and Allocation:** This component allocates virtual machines to the Cloud nodes based on user's QoS targets and the Clouds energy management goals. On receiving a user application, the scheduler does the following: (i) consults the Application Composition Engine about availability of software and hardware infrastructure services that are required to satisfy the request locally, (ii) asks the Sensor component to submit feedback on the local Cloud nodes' energy consumption and utilization status; and (iii) enquires the Market and Policy Engine about accountability of the submitted request. A request is termed as accountable if the concerning user has available credits in the Cloud bank and based on the specified QoS constraints the establishment of SLA is feasible. In case all three components reply favorably, the application is hosted locally and is periodically monitored until it finishes execution.

Data center resources may deliver different levels of performance to their clients; hence, QoS-aware resource selection plays an important role in Cloud computing. Additionally, Cloud applications can present varying workloads. It is therefore essential to carry out a study of Cloud services and their workloads in order to identify common behaviors, patterns, and explore load forecasting approaches that can potentially lead to more efficient scheduling and allocation. In this context, there is need to analyse sample applications and correlations between workloads, and attempt to build performance models that can help explore trade-offs between QoS targets.

**Market and Policy Engine:** The SLA module stores the service terms and conditions that are being supported by the Cloud to each respective Cloud Broker on a per user basis. Based on these terms and conditions, the Pricing module can determine how service requests are charged based on the available supply and required demand of computing resources within the Cloud. The Accounting module stores the actual usage information of resources by requests so that the total usage cost of each user can be calculated. The Billing module then charges the usage costs to users accordingly.

Cloud customers can normally associate two or more conflicting QoS targets with their application services. In such cases, it is necessary to trade off one or more QoS targets to find a superior solution. Due to such diverse QoS targets and varying optimization objectives, we end up with a Multi-dimensional Optimization Problem (MOP). For solving the MOP, one can explore multiple heterogeneous optimization algorithms, such as dynamic programming, hill climbing, parallel swarm optimization, and multi-objective genetic algorithm.

**Application Composition engine:** This component of the Cloud Coordinator encompasses a set of features intended to help application developers create and deploy [18] applications, including the ability for on demand interaction with a database backend such as SQL Data services provided by Microsoft Azure, an application server such as Internet Information Server (IIS) enabled with secure ASP.Net scripting engine to host web applications, and a SOAP driven Web services API for programmatic access along with combination and integration with other applications and data.

**Virtualization:** VMs support flexible and utility driven configurations that control the share of processing power they can consume based on the time criticality of the underlying application. However, the current approaches to VM-based Cloud computing are limited to rather inflexible configurations within a Cloud. This limitation can be solved by developing mechanisms for transparent migration of VMs across service boundaries with the aim of minimizing cost of service delivery (e.g., by migrating to a Cloud located in a region where the energy cost is low) and while still meeting the SLAs. The Mobility Manager is responsible for dynamic migration of VMs based on the real-time feedback given by the Sensor service. Currently, hypervisors such as VMware [8] and Xen [9] have a limitation that VMs can only be migrated between hypervisors that are within the same sub-net and share common storage. Clearly, this is a serious bottleneck to achieve adaptive migration of VMs in federated Cloud environments. This limitation has to be addressed in order to support utility driven, power-aware migration of VMs across service domains.

**Sensor**: Sensor infrastructure will monitor the power consumption, heat dissipation, and utilization of computing nodes in a virtualized Cloud environment. To this end, we will extend our Service Oriented Sensor Web [14] software system. Sensor Web provides a middleware infrastructure and programming model for creating, accessing, and utilizing tiny sensor devices that are deployed within a Cloud. The Cloud Coordinator service makes use of Sensor Web services for dynamic sensing of Cloud nodes and surrounding temperature. The output data reported by sensors are feedback to the Coordinator's Virtualization and Scheduling components, to optimize the placement, migration, and allocation of VMs in the Cloud. Such sensor-based real time monitoring of the Cloud operating environment aids in avoiding server breakdown and achieving optimal throughput out of the available computing and storage nodes.

**Discovery and Monitoring:** In order to dynamically perform scheduling, resource allocation, and VM migration to meet SLAs in a federated network, it is mandatory that up-to-date information related to Cloud's availability, pricing and SLA rules are made available to the outside domains via the Cloud Exchange. This component of Cloud Coordinator is solely responsible for interacting with the Cloud Exchange through remote messaging. The Discovery and Monitoring component undertakes the following activities: (i) updates the resource status metrics including utilization, heat dissipation, power consumption based on feedback given by the Sensor component; (ii) facilitates the Market and Policy Engine in periodically publishing the pricing policies, SLA rules to the Cloud Exchange; (iii) aids the Scheduling and Allocation component in dynamically discovering the Clouds that offer better optimization for SLA constraints such as deadline and budget limits; and (iv) helps the Virtualization component in determining load and power consumption; such information aids the Virtualization component in performing load-balancing through dynamic VM migration.

Further, system components will need to share scalable methods for collecting and representing monitored data. This leads us to believe that we should interconnect and monitor system components based on decentralized messaging and information indexing infrastructure, called Distributed Hash Tables (DHTs) [26]. However, implementing scalable techniques that monitor the dynamic behaviors related to services and resources is non-trivial. In order to support a scalable service monitoring algorithm over a DHT infrastructure, additional data distribution indexing techniques such as logical multi-dimensional or spatial indices [27] (MX-CIF Quad tree, Hilbert Curves, Z Curves) should be implemented.

### 3.2 Cloud Broker (CB)

The Cloud Broker acting on behalf of users identifies suitable Cloud service providers through the Cloud Exchange and negotiates with Cloud Coordinators for an allocation of resources that meets QoS needs of users. The architecture of Cloud Broker is shown in Figure 3 and its components are discussed below:

**User Interface:** This provides the access linkage between a user application interface and the broker. The Application Interpreter translates the execution requirements of a user application which include what is to be executed, the description of task inputs including remote data files (if required), the information about task outputs (if present), and the desired QoS. The Service Interpreter understands the service requirements needed for the execution which comprise service location, service type, and specific details such as remote batch job submission systems for computational services. The Credential Interpreter reads the credentials for accessing necessary services.

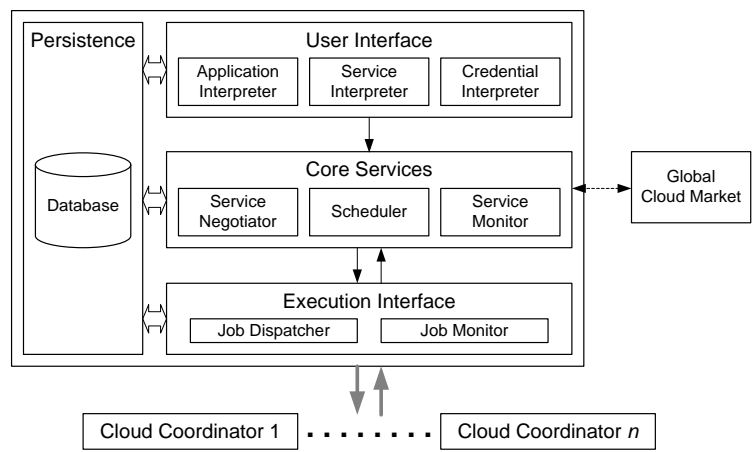

Figure 3: High level architecture of Cloud Broker service.

**Core Services:** They enable the main functionality of the broker. The Service Negotiator bargains for Cloud services from the Cloud Exchange. The Scheduler determines the most appropriate Cloud services for the user application based on its application and service requirements. The Service Monitor maintains the status of Cloud services by periodically checking the availability of known Cloud services and discovering new services that are available. If the local Cloud is unable to satisfy application requirements, a Cloud Broker lookup request that encapsulates the user's QoS parameter is submitted to the Cloud Exchange, which matches the lookup request against the available offers. The matching procedure considers two main system performance metrics: first, the user specified QoS targets must be satisfied within acceptable bounds and, second, the allocation should not lead to overloading (in terms of utilization, power consumption) of the nodes. In case the match occurs the quote is forwarded to the requester (Scheduler). Following that, the Scheduling and Allocation component deploys the application with the Cloud that was suggested by Cloud market.

**Execution Interface:** This provides execution support for the user application. The Job Dispatcher creates the necessary broker agent and requests data files (if any) to be dispatched with the user application to the remote Cloud resources for execution. The Job Monitor observes the execution status of the job so that the re-sults of the job are returned to the user upon job completion.

**Persistence:** This maintains the state of the User Interface, Core Services, and Execution Interface in a database. This facilitates recovery when the broker fails and assists in user-level accounting.

### 3.3 Cloud Exchange (CEx)

As a market maker, the CEx acts as an information registry that stores the Cloud's current usage costs and demand patterns. Cloud Coordinators periodically update their availability, pricing, and SLA policies with the CEx. Cloud Brokers query the registry to learn information about existing SLA offers and resource availability of member Clouds in the federation. Furthermore, it provides match-making services that map user requests to suitable service providers. Mapping functions will be implemented by leveraging various economic models such as Continuous Double Auction (CDA) as proposed in earlier works [6].

As a market maker, the Cloud Exchange provides directory, dynamic bidding based service clearance, and payment management services as discussed below.

- **Directory:** The market directory allows the global CEx participants to locate providers or consumers with the appropriate bids/offers. Cloud providers can publish the available supply of resources and their offered prices. Cloud consumers can then search for suitable providers and submit their bids for required resources. Standard interfaces need to be provided so that both providers and consumers can access resource information from one another readily and seamlessly.

- **Auctioneer:** Auctioneers periodically clear bids and asks received from the global CEx participants. Auctioneers are third party controllers that do not represent any providers or consumers. Since the auctioneers are in total control of the entire trading process, they need to be trusted by participants.

- **Bank:** The banking system enforces the financial transactions pertaining to agreements between the global CEx participants. The banks are also independent and not controlled by any providers and consumers; thus facilitating impartiality and trust among all Cloud market participants that the financial transactions are conducted correctly without any bias. This should be realized by integrating with online payment management services, such as PayPal, with Clouds providing accounting services.

## 4. Early Experiments and Preliminary Results

Although we have been working towards the implementation of a software system for federation of cloud computing environments, it is still a work-in-progress. Hence, in this section, we present our experiments and evaluation that we undertook using CloudSim [29] framework for studying the feasibility of the proposed research vision. The experiments were conducted on a Celeron machine having the following configuration: 1.86GHz with 1MB of L2 cache and 1 GB of RAM running a standard Ubuntu Linux version 8.04 and JDK 1.6.

### 4.1. Evaluating Performance of Federated Cloud Computing Environments

The first experiment aims at proving that federated infrastructure of clouds has potential to deliver better performance and service quality as compared to existing non-federated approaches. To this end, a simulation environment that models federation of three Cloud providers and a user (Cloud Broker) is modeled. Every provider instantiates a Sensor component, which is responsible for dynamically sensing the availability information related to the local hosts. Next, the sensed statistics are reported to the Cloud Coordinator that utilizes the information in undertaking load-migration decisions. We evaluate a straightforward load-migration policy that performs online migration of VMs among federated Cloud providers only if the origin provider does not have the requested number of free VM slots available. The migration process involves the following steps: (i) creating a virtual machine instance that has the same configuration, which is supported at the destination provider; and (ii) migrating the applications assigned to the original virtual machine to the newly instantiated virtual machine at the destination provider. The federated network of Cloud providers is created based on the topology shown in Figure 4.

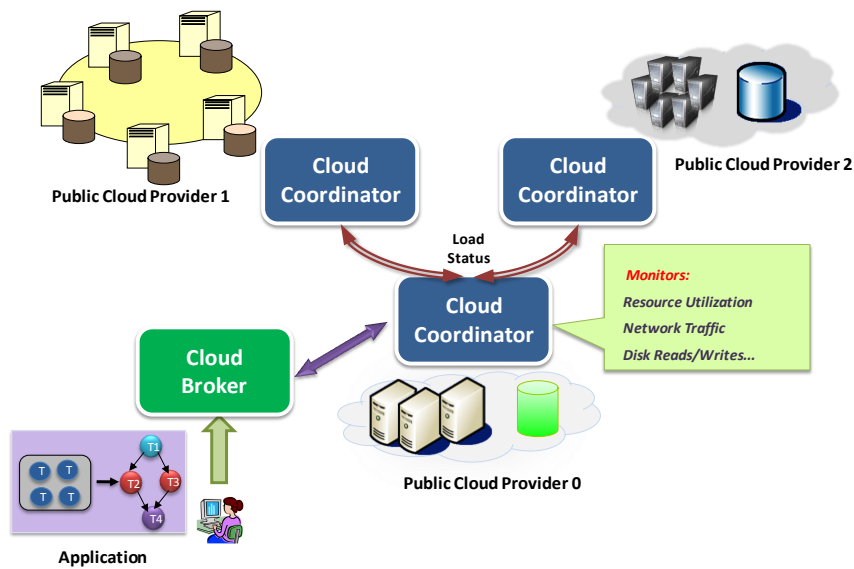

Figure 4: A network topology of federated Data Centers.

Every Public Cloud provider in the system is modeled to have 50 computing hosts, 10GB of memory, 2TB of storage, 1 processor with 1000 MIPS of capacity, and a time-shared VM scheduler. Cloud Broker on behalf of the user requests instantiation of a VM that requires 256MB of memory, 1GB of storage, 1 CPU, and time-shared Cloudlet scheduler. The broker requests instantiation of 25 VMs and associates one Cloudlet (Cloud application abstraction) to each VM to be executed. These requests are originally submitted with the Cloud Provider 0. Each

Cloudlet is modeled to have 1800000 MIs (Million Instrictions). The simulation experiments were run under the following system configurations: (i) a federated network of clouds is available, hence data centers are able to cope with peak demands by migrating the excess of load to the least loaded ones; and (ii) the data centers are modeled as independent entities (without federation). All the workload submitted to a Cloud provider must be processed and executed locally.

Table 2 shows the average turn-around time for each Cloudlet and the overall makespan of the user application for both cases. A user application consists of one or more Cloudlets with sequential dependencies. The simulation results reveal that the availability of federated infrastructure of clouds reduces the average turn-around time by more than 50%, while improving the makespan by 20%. It shows that, even for a very simple load-migration policy, availability of federation brings significant benefits to user's application performance.

Table 2: Performance Results.

| Performance Metrics | With Federation | Without Federation | % Improvement |
|---|---|---|---|
| Average Turn Around Time (Secs) | 2221.13 | 4700.1 | > 50% |
| Makespan (Secs) | 6613.1 | 8405 | 20% |

### 4.2 Evaluating a Cloud provisioning strategy in a federated environment

In previous subsection, we focused on evaluation of the federated service and resource provisioning scenarios. In this section, a more complete experiment that also models the inter-connection network between federated clouds, is presented. This example shows how the adoption of federated clouds can improve productivity of a company with expansion of private cloud capacity by dynamically leasing resources from public clouds at a reasonably low cost.

The simulation scenario is based on federating a private cloud with the Amazon EC2 cloud. The public and the private clouds are represented as two data centers in the simulation. A Cloud Coordinator in the private data center receives the user's applications and processes them in a FCFS basis, queuing the tasks when there is available capacity for them in the infrastructure. To evaluate the effectiveness of a hybrid cloud in speeding up tasks execution, two scenarios are simulated. In the first scenario, tasks are kept in the waiting queue until active tasks finish (currently executing) in the private cloud. All the workload is processed locally within the private cloud. In the second scenario, the waiting tasks are directly sent to available public cloud. In other words, second scenario simulates a Cloud Bursts case for integrating local private cloud with public cloud form handing peak in service demands. Before submitting tasks to the Amazon cloud, the VM images (AMI) are loaded and instantiated. The number of images instantiated in the Cloud is varied in the experiment, from 10% to 100% of the number of machines available in the private cloud. Once images are created, tasks in the waiting queues are submitted to them, in such a way that only one task run on each VM at a given instance of time. Every time a task finishes, the next task in the waiting queue is submitted to the available VM host. When there were no tasks to be submitted to the VM, it is destroyed in the cloud.

The local private data center hosted 100 machines. Each machine has 2GB of RAM, 10TB of storage and one CPU run 1000 MIPS. The virtual machines created in the public cloud were based in an Amazon's small instance (1.7 GB of memory, 1 virtual core, and 160 GB of instance storage). We consider in this example that the virtual core of a small instance has the same processing power as the local machine.

The workload sent to the private cloud is composed of 10000 tasks. Each task takes between 20 and 22 minutes to run in one CPU. The exact amount of time was randomly generated based on the normal distribution. Each of the 10000 tasks is submitted at the same time to the scheduler queue.

Table 3 shows the makespan of the tasks running only in the private cloud and with extra allocation of resources from the public cloud. In the third column, we quantify the overall cost of the services. The pricing policy was designed based on the Amazon's small instances (U\$ 0.10 per instance per hour) pricing model. It means that the cost per instance is charged hourly in the beginning of execution. And, if an instance runs during 1 hour and 1 minute, the amount for 2 hours (U\$ 0.20) will be charged.

Table 3: Cost and performance of several public/private cloud strategies

|  | Makespan (s) | Cloud Cost (U\$) |
|---|---|---|
| Private only | 127155.77 | 0.00 |
| Public 10% | 115902.34 | 32.60 |
| Public 20% | 106222.71 | 60.00 |
| Public 30% | 98195.57 | 83.30 |
| Public 40% | 91088.37 | 103.30 |
| Public 50% | 85136.78 | 120.00 |
| Public 60% | 79776.93 | 134.60 |
| Public 70% | 75195.84 | 147.00 |
| Public 80% | 70967.24 | 160.00 |
| Public 90% | 67238.07 | 171.00 |
| Public 100% | 64192.89 | 180.00 |

Increasing the number of resources by a rate reduces the job makespan at the same rate, which is an expected observation or outcome. However, the cost associated with the processing increases significantly at higher rates. Nevertheless, the cost is still acceptable, considering that peak demands happen only occasionally and that most part of time this extra public cloud capacity is not required. So, leasing public cloud resources is cheapest than buying and maintaining extra resources that will spend most part of time idle.

## 5. Conclusions and Future Directions

Development of fundamental techniques and software systems that integrate distributed clouds in a federated fashion is critical to enabling composition and deployment of elastic application services. We believe that outcomes of this research vision will make significant scientific advancement in understanding the theoretical and practical problems of engineering services for federated environments. The resulting framework facilitates the federated management of system components and protects customers with guaranteed quality of services in large, federated and highly dynamic environments. The different components of the proposed framework offer powerful capabilities to address both services and resources management, but their end-to-end combination aims to dramatically improve the effective usage, management, and administration of Cloud systems. This will provide enhanced degrees of scalability, flexibility, and simplicity for management and delivery of services in federation of clouds.

In our future work, we will focus on developing comprehensive model driven approach to provisioning and delivering services in federated environments. These models will be important because they allow adaptive system management by establishing useful relationships between high-level performance targets (specified by operators) and low-level control parameters and observables that system components can control or monitor. We will model the behaviour and performance of different types of services and resources to adaptively transform service requests. We will use a broad range of analytical models and statistical curve-fitting techniques such as multi-class queuing models and linear regression time series. These models will drive and possibly transform the input to a service provisioner, which improves the efficiency of the system. Such improvements will better ensure the achievement of performance targets, while reducing costs due to improved utilization of resources. It will be a major advancement in the field to develop a robust and scalable system monitoring infrastructure to collect real-time data and readjust these models dynamically with a minimum of data and training time. We believe that these models and techniques are critical for the design of stochastic provisioning algorithms across large federated Cloud systems where resource availability is uncertain.

Lowering the energy usage of data centers is a challenging and complex issue because computing applications and data are growing so quickly that increasingly larger servers and disks are needed to process them fast enough within the required time period. Green Cloud computing is envisioned to achieve not only efficient processing and utilization of computing infrastructure, but also minimization of energy consumption. This is essential for ensuring that the future growth of Cloud Computing is sustainable. Otherwise, Cloud computing with increasingly pervasive front-end client devices interacting with back-end data centers will cause an enormous escalation of energy usage. To address this problem, data center resources need to be managed in an energy-efficient manner to drive Green Cloud computing. In particular, Cloud resources need to be allocated not only to satisfy QoS targets specified by users via Service Level

Agreements (SLAs), but also to reduce energy usage. This can be achieved by applying market-based utility models to accept requests that can be fulfilled out of the many competing requests so that revenue can be optimized along with energy-efficient utilization of Cloud infrastructure.

**Acknowledgements:** We acknowledge all members of Melbourne CLOUDS Lab (especially William Voorsluys and Suraj Pandey) for their contributions to InterCloud investigation.